\documentclass{amsart}
\usepackage{amssymb}
\usepackage{amsfonts}
\usepackage{amsmath}
\usepackage{graphicx}
\usepackage{epstopdf}
\usepackage[backref]{hyperref}

\begin{document}

	\begin{center}
		{\Large \bf 
		Two-fold refinement of non simply laced Chern-Simons theories\\
			\vspace*{1 cm}
			
			{\large  M.Y. Avetisyan and R.L.Mkrtchyan
			}
			\vspace*{0.2 cm}
			
			{\small\it Yerevan Physics Institute, 2 Alikhanian Br. Str., 0036 Yerevan, Armenia}
			
		}
		
	\end{center}\vspace{2cm}

{\small  {\bf Abstract.} 

Inspired by the two-parameter Macdonald-Cherednik deformation of the formulae for non simply laced simple Lie algebras, 
we propose a two-fold refinement of the partition function
of the corresponding Chern-Simons theory on $S^3$. It is based on a two-fold refinement of the Kac-Peterson formula for the volume of the fundamental domain of 
the coroot lattice of a non simply laced Lie algebras. 
We further derive explicit integral representations of the two-fold refined Chern-Simons partition functions. We also present the corresponding generalized universal-like expressions for 
them. With these formulae in hand one can try to investigate a possible duality of the corresponding Chern-Simons theories with
hypothetical two-fold refined topological string theories.


\section{Introduction}

In \cite{AM21} we presented an explicit partition function of refined Chern-Simons theory on $S^3$ for all gauge simple Lie algebras. It coincides with Aganagich-Shakirov \cite{AS11,AS12a} and Aganagich-Schaeffer \cite{AS12} partition functions for A and D algebras, as well as with their non-perturbative form, given in \cite{k_s,KM}. In the non-refined limit 
it coincided with the  universal form \cite{MV,M13} of Witten's partition function \cite{W1} for all simple gauge algebras. The refinement of Chern-Simons theory is based on the Macdonald's deformation of a number of formulae in the theory of simple Lie algebras, proved later in Cherednik's \cite{Cher1,Cher2}, and others. 
In \cite{Cher1,Cher2} it is shown, that for non simply laced algebras Macdonalds deformation can have two parameters, corresponding to two different lengths of the roots of the algebras.
 The aim of the present paper is to generalize the partition function \cite{AM21} of the refined Chern-Simons theory into the two-fold refined one, adding the second deformation parameter corresponding to that of Cherednik. 

The main aim of \cite{AM21} was the preparatory work for finding the dual refined topological string  for refined Chern-Simons theory with an arbitrary simple gauge algebra. That goal was achieved for all classical simple gauge algebras in the next paper \cite{AM22}, which required lengthy calculations. Similar calculations for the two-fold refined theories, considered in the present work, evidently would be much more complex, so we do not carry out them in the present work. However, they will hopefully establish the two-parameter deformation of topological string theories, unknown at the moment. 

Our two-fold refined version of the Chern-Simons partition function is presented in Section \ref{partfunc} and is based on the generalization of Kac-Peterson formula for the determinant of the symmetrized Cartan matrix \cite{KP}. This formula was already generalized in \cite{AM21} to include one refinement parameter. In the present paper we present its two-fold refined version. In Section \ref{intrep} we present an integral representation of the partition function, which includes the non-perturbative (w.r.t. the string coupling constant) corrections, as shown earlier in the single-refined case in \cite{KM}. 

In the Section \ref{sect:univ} we present this two-fold refined partition function in a "universal" form for all non simply laced algebras, generalizing the corresponding expression from \cite{AM21}. This form, as was mentioned, is ready for the further transformation of the partition function into the form, corresponding to the (hypothetical two-fold refined) topological strings.

\section{Double refinement of Kac-Peterson identity and the corresponding Chern-Simons partition functions} \label{partfunc}

In \cite{AM21} we introduced a deformation (refinement) of the Kac-Peterson formula for the volume of the fundamental domain of the coroots lattice \cite{KP}
via the inclusion of a parameter $y$. 
It is significant for the Chern-Simons theory since that volume is a part of the partition function of the Chern-Simons theory on a 3d sphere $S^3$, see below.  The refined formula looks as follows:

\begin{eqnarray} \label{vol}
	Vol(Q^{\vee})=  (ty)^{-\frac{r}{2}}  \prod_{m=0}^{y-1} \prod_{\alpha_+} 2\sin \pi \frac{y(\alpha,\rho)-m (\alpha,\alpha)/2}{ty}
\end{eqnarray}
where $Q^{\vee}$ is the coroots lattice, $Vol(Q^{\vee})$ is the volume of its fundamental domain, $r$ is the rank of algebra, $\alpha_+$ are its positive roots, $\rho$ is the half-sum of 
the positive roots, $t$ is the dual Coxeter number, taken in an arbitrary normalization of the invariant Cartan-Killing metric (or, the same, the sum of Vogel's universal parameters,
 see e.g. \cite{AM21}). Here the refinement parameter $y$ is assumed to be a positive integer. At $y=1$ this formula coincides with the original formula of Kac-Peterson, see 
\cite{KP}, eq. (4.32.2).

It appears, that the further generalization of this formula is possible, with inclusion of two parameters: 

\begin{eqnarray} \label{vol2}
	Vol(Q^{\vee})=  (\tilde{k})^{-\frac{r}{2}}  \prod_{\alpha_+}  \prod_{m=0}^{k_{\nu_\alpha}-1} 2\sin \pi \frac{k_s(\rho_s,\alpha)+k_l(\rho_l,\alpha)-m (\alpha,\alpha)/2}{	\tilde{k}}
\end{eqnarray}

 \begin{eqnarray}
	\tilde{k}=k_s(\rho_s,\theta)+k_l(\rho_l,\theta)+k_l\frac{(\theta,\theta)}{2}
\end{eqnarray}

where $k_l$ and $k_s$ are the  refinement parameters, here taken to be positive integers, corresponding to the different lengths of the roots, with subscripts $s$ for the short and $l$ for the long ones.
$\rho_s$ and $\rho_l$ stand for the half-sum of all positive short and long roots, correspondingly. The subscript $\nu_\alpha$ is $s$ if $\alpha$ is a short root, and $l$ if $\alpha$ is
a long one. $\theta$ stands for the highest root of an algebra.

Note, that this generalization touches the non simply laced algebras only. The previous case with one deformation parameter $y$ is covered in $k_s=k_l=y$ case.

This formula is checked numerically for all types of non simply laced simple Lie algebras for dozen thousands of random values of ranks and parameters $k_s, k_l$, so we assume it 
is correct in all cases. 
The analytical proof of this equality perhaps can be carried out similarly as in \cite{AM21}, making use of the following identity:

\begin{eqnarray}
	N=\prod_{k=1}^{N-1}2 \sin{\pi\frac{k}{N}}
\end{eqnarray}

The partition function of the refined Chern-Simons theory on $S^3$ for all gauge algebras, suggested in \cite{AM21}, is 

 \begin{eqnarray}\label{refCS}
	Z(\kappa,y)= Vol(Q^{\vee})^{-1} \delta^{-\frac{r}{2}} \prod_{m=0}^{y-1} \prod_{\alpha_+} 2\sin \pi \frac{y(\alpha,\rho)-m (\alpha,\alpha)/2}{\delta}
\end{eqnarray}
 where $\delta=\kappa+yt$, $\kappa$ is the coupling constant.  At $y=1$ this coincides with partition function of unrefined Chern-Simons theory, derived in pioneering \cite{W1}. It has the following 
 key feature 
 
 \begin{eqnarray}
 	Z(0,y)=1
 \end{eqnarray}
due to the (\ref{vol}) identity. The meaning of this identity is as follows: a Chern-Simons theory with the coupling constant $\kappa$
 (which is a positive integer in the Cartan-Killing normalization of the scalar 
product in the root space) is based on the unitary representations of level $\kappa$ which are of a finite number. At $\kappa=0$ there is only one – trivial representation.

Generalizing the partition function (\ref{refCS}) we suggest the following partition function for the two-fold refined CS on $S^3$:

\begin{eqnarray}\label{Z2ref1}
Z(\kappa,k_s,k_l)= Vol(Q^{\vee})^{-1} \delta^{-\frac{r}{2}} 	 \prod_{\alpha_+} \prod_{m=0}^{k_{\nu_\alpha}-1} \sin \pi \frac{{k_s(\rho_s,\alpha)+k_l(\rho_l,\alpha)-m (\alpha,\alpha)/2}}{\delta}
\end{eqnarray}
with $\delta=\kappa+\tilde{k}$. 

This partition function also satisfies the key identity 

\begin{eqnarray}\label{key2}
	Z(0,k_s,k_l)=1
\end{eqnarray}
due to the generalized identity (\ref{vol2}).

Making use of the key identity (\ref{key2}), similarly to \cite{AM21}, one can easily transform (\ref{Z2ref1}) into the following form

\begin{eqnarray} \label{Z2ref}
	Z(\kappa,k_s,k_l)=  (\frac{\tilde{k}}{\tilde{k}+\kappa})^{\frac{r}{2}} \prod_{\alpha_+} \prod_{m=0}^{k_{\nu_\alpha}-1} \frac{\sin \pi \frac{{k_s(\rho_s,\alpha)+k_l(\rho_l,\alpha)-m (\alpha,\alpha)/2}}{\tilde{k}+\kappa}}
	{\sin \pi \frac{k_s(\rho_s,\alpha)+k_l(\rho_l,\alpha)-m (\alpha,\alpha)/2}{\tilde{k}}}
\end{eqnarray}
which is convenient for further transformation into universality-like integral representation of \cite{AM21}, appropriate for a further duality establishment \cite{AM22}.

\section{Integral representation of the partition function for the two-fold refined CS theories} \label{intrep}

We rewrite (\ref{Z2ref})  into the separate products over the short and the long roots:
 
 \begin{multline}
	Z(\kappa,k_s,k_l)= (\frac{\tilde{k}}{\tilde{k}+\kappa})^{\frac{r}{2}+(k_s L_s+k_l L_l)}  \\
	\prod_{\alpha_+^s} \prod_{m=0}^{k_s-1} 
	\frac{\sin \pi \frac{{k_s(\rho_s,\alpha)+k_l(\rho_l,\alpha)-m (\alpha,\alpha)/2}}{\tilde{k}+\kappa}}
	{ \pi \frac{k_s(\rho_s,\alpha)+k_l(\rho_l,\alpha)-m (\alpha,\alpha)/2}{\tilde{k}+\kappa}} \cdot
	\frac{ \pi \frac{k_s(\rho_s,\alpha)+k_l(\rho_l,\alpha)-m (\alpha,\alpha)/2}{\tilde{k}}}
	{\sin \pi \frac{{k_s(\rho_s,\alpha)+k_l(\rho_l,\alpha)-m (\alpha,\alpha)/2}}{\tilde{k}}} \times \\
	\prod_{\alpha_+^l} \prod_{m=0}^{k_l-1} 
	\frac{\sin \pi \frac{{k_s(\rho_s,\alpha)+k_l(\rho_l,\alpha)-m (\alpha,\alpha)/2}}{\tilde{k}+\kappa}}
	{ \pi \frac{k_s(\rho_s,\alpha)+k_l(\rho_l,\alpha)-m (\alpha,\alpha)/2}{\tilde{k}+\kappa}} \cdot 
	\frac{ \pi \frac{k_s(\rho_s,\alpha)+k_l(\rho_l,\alpha)-m (\alpha,\alpha)/2}{\tilde{k}}}
	{\sin \pi \frac{{k_s(\rho_s,\alpha)+k_l(\rho_l,\alpha)-m (\alpha,\alpha)/2}}{\tilde{k}}}= \\
	  (\frac{\tilde{k}}{\tilde{k}+\kappa})^{\frac{r}{2}+(k_s L_s+k_l L_l)} 
	Z_1^s\cdot Z_2^s\cdot Z_1^l\cdot Z_2^l
\end{multline}

  Here $L_s$ and $L_l$ denote the number of positive short and long roots, respectively. Last equality implies a natural definition of $Z_1^s, Z_2^s, Z_1^l, Z_2^l$.
  
 Using the well-known identity
 \begin{eqnarray}
	\frac{\sin \pi z}{\pi z} = \frac{1}{\Gamma(1+z) \Gamma(1-z)}
\end{eqnarray}

and the following integral representation

\begin{eqnarray}
	\ln\Gamma(1+z)=\int_{0}^{\infty}dx \frac{e^{-zx}+z(1-e^{-x})-1}{x(e^{x}-1)}
\end{eqnarray}

we have

\begin{eqnarray*}
\ln Z_1^s=	-\int_{0}^{\infty} \frac{dx}{x(e^{x}-1)} \sum_{m=0}^{k_s-1}\sum_{\alpha_+^s} ( e^{-x\frac{k_s(\rho_s,\alpha)+k_l(\rho_l,\alpha)-m (\alpha,\alpha)/2}{\tilde{k}+\kappa}}+
	e^{x\frac{k_s(\rho_s,\alpha)+k_l(\rho_l,\alpha)-m (\alpha,\alpha)/2}{\tilde{k}+\kappa}}-2)
\end{eqnarray*}

and

\begin{eqnarray*}
\ln Z_2^s=	\int_{0}^{\infty} \frac{dx}{x(e^{x}-1)} \sum_{m=0}^{k_s-1}\sum_{\alpha_+^s} ( e^{-x\frac{k_s(\rho_s,\alpha)+k_l(\rho_l,\alpha)-m (\alpha,\alpha)/2}{\tilde{k}}}+
	e^{x\frac{k_s(\rho_s,\alpha)+k_l(\rho_l,\alpha)-m (\alpha,\alpha)/2}{\tilde{k}}}-2)
\end{eqnarray*}

Similarly,

\begin{eqnarray*}
\ln Z_1^l=	-\int_{0}^{\infty} \frac{dx}{x(e^{x}-1)} \sum_{m=0}^{k_l-1}\sum_{\alpha_+^l} ( e^{-x\frac{k_s(\rho_s,\alpha)+k_l(\rho_l,\alpha)-m (\alpha,\alpha)/2}{\tilde{k}+\kappa}}+
	e^{x\frac{k_s(\rho_s,\alpha)+k_l(\rho_l,\alpha)-m (\alpha,\alpha)/2}{\tilde{k}+\kappa}}-2)
\end{eqnarray*}

and

\begin{eqnarray*}
\ln Z_2^l=	\int_{0}^{\infty} \frac{dx}{x(e^{x}-1)} \sum_{m=0}^{k_l-1}\sum_{\alpha_+^l} ( e^{-x\frac{k_s(\rho_s,\alpha)+k_l(\rho_l,\alpha)-m (\alpha,\alpha)/2}{\tilde{k}}}+
	e^{x\frac{k_s(\rho_s,\alpha)+k_l(\rho_l,\alpha)-m (\alpha,\alpha)/2}{\tilde{k}}}-2)
\end{eqnarray*}

Let us compute $\ln Z_1^s+\ln Z_1^l$ and $\ln Z_2^s+\ln Z_2^l$:

\begin{eqnarray*}
\ln Z_1:=\ln Z_1^s+\ln Z_1^l=-\int_{0}^{\infty} \frac{dx}{x(e^{x}-1)} 
\big(F_X(\frac{x}{\tilde{k}+\kappa},k_s,k_l)-r-2(k_s L_s+k_l L_l)\big), \\
\ln Z_2:=\ln Z_2^s+\ln Z_2^l=\int_{0}^{\infty} \frac{dx}{x(e^{x}-1)} 
\big(F_X(\frac{x}{\tilde{k}},k_s,k_l)-r-2(k_s L_s+k_l L_l)\big)=\\
\int_{0}^{\infty} \frac{dx}{x(e^{x\frac{\tilde{k}}{\tilde{k}+\kappa}}-1)} 
\big(F_X(\frac{x}{\tilde{k}+\kappa},k_s,k_l)-r-2(k_s L_s+k_l L_l)\big)
\end{eqnarray*}

where

\begin{eqnarray}\label{FX}
		F_X(x,k_s,k_l)=  \\ \nonumber
		  r+	\sum_{m=0}^{k_{\nu_\alpha}-1}	\sum_{\alpha_{+}}  ( 
		e^{x(k_s(\rho_s,\alpha)+k_l(\rho_l,\alpha)-m (\alpha,\alpha)/2)}+e^{-x(k_s(\rho_s,\alpha)+k_l(\rho_l,\alpha)-m (\alpha,\alpha)/2)})
\end{eqnarray}

Using the relation
\begin{eqnarray} \label{cotmcotId}
	\frac{1}{e^{b x}-1}-\frac{1}{e^{a x} -1} = \frac{e^{a x}-e^{b x}}{(e^{a x}-1)(e^{b x }-1)}=\frac{\sinh (\frac{x(a-b)}{2})}{2\sinh (\frac{x a}{2})\sinh (\frac{x b}{2})}\,,
\end{eqnarray}

we have:
\begin{eqnarray*}
\ln Z_1+\ln Z_2=
\int_{0}^{\infty} \frac{dx}{2x} \frac{sh(\frac{x}{2}\frac{\kappa}{\tilde{k}+\kappa})}
{sh(\frac{x}{2}) sh(\frac{x}{2}\frac{\tilde{k}}{\tilde{k}+\kappa})}
\big(F_X(\frac{x}{\tilde{k}+\kappa},k_s,k_l)-r-2(k_s L_s+k_l L_l)\big)= \\
\int_{0}^{\infty} \frac{dx}{2x} \frac{sh(x\kappa)}
{sh(x(\tilde{k}+\kappa)) sh(x\tilde{k})}
\big(F_X(2x,k_s,k_l)-r-2(k_s L_s+k_l L_l)\big)=\\
\frac{1}{4} \int_{R_+} \frac{dx}{x} \frac{sh(x\kappa)}
{sh(x(\tilde{k}+\kappa)) sh(x\tilde{k})}
\big(F_X(2x,k_s,k_l)-r-2(k_s L_s+k_l L_l)\big)
\end{eqnarray*}
where in the last line we firstly used that the integrand is even under $x\rightarrow -x$ transformation and secondly that its residue is zero (for the same reason), 
so we can slightly deform the integration contour into a small semicircle in the upper semiplane around the zero (that is the definition of a contour $R_+$). 
This form will be used below.

Now we can write the overall integral representation for $\ln Z$:
\begin{eqnarray*}
\ln Z=(\frac{r}{2}+(k_s L_s+k_l L_l))\cdot \ln  (\frac{\tilde{k}}{\tilde{k}+\kappa})+ \\
\frac{1}{4} \int_{R_+} \frac{dx}{x} \frac{sh(x\kappa)}
{sh(x(\tilde{k}+\kappa)) sh(x\tilde{k})}
\big(F_X(2x,k_s,k_l)-r-2(k_s L_s+k_l L_l)\big)
\end{eqnarray*}

Due to the following identity
\begin{eqnarray}
	\frac{1}{4}\int_{R_+} \frac{dx}{x} \frac{\sinh (x(a-b))}{\sinh (x a)\sinh (x b)}=-\frac{1}{2}\log (\frac{a}{b})\,,
\end{eqnarray}

we finally have:
\begin{eqnarray*}
\ln Z=\frac{1}{4} \int_{R_+} \frac{dx}{x} \frac{sh(x\kappa)}
{sh(x(\tilde{k}+\kappa)) sh(x\tilde{k})}
\big(F_X(2x,k_s,k_l)\big)
\end{eqnarray*}

thus obtaining the final expression for the partition function of the two-fold refined Chern-Simons theory on the $S^3$ manifold.

All formulae of course coincide with those for the usual refined Chern-Simons in the case $k_s=k_l$. Particularly $F_X(x,k,k)=F_X(x,k)$ with the last function defined in \cite{AM21}

\section{The universal-like expressions for the two-fold refined CS for non simply laced algebras} \label{sect:univ}

In this section we calculate the function $ F_X(x,k_s,k_l) $ in a form which naturally extends it into arbitrary values of the refinement parameters $k_s, k_l$. It is also 
convenient for a further establishment of duality with the topological strings. 

We will show that for all non simply laced algebras one can present the $F_X(x,k_s,k_l)$ in the form of $\frac{A_{X}(k_s,k_l)}{B_{X}(k_s,k_l)}$, given below. 

Let us consider the $B_n$ algebras. Normalization corresponds to $\alpha=-4$, i.e. the square of the long root is $4$.
 The corresponding representation we mentioned above is

\begin{eqnarray}\label{F2}
	F_{B_n}(x,k_s,k_l)= 
	\frac{A_{B_n}(k_s,k_l)}{B_{B_n}(k_s,k_l)} \\
	B_{B_n(k_s,k_l)}= (q^2-1)  (q^{4 k_l}-1)\\
	A_{B_n(k_s,k_l)}=\\
	 (q^{2  k_l n}-1) q^{-2 (2 k_l n+k_l+ k_s)}\times\\
	  (-q^{4   k_l (n+1)+2   k_s+1}+q^{2 k_l (n+2)+2 k_s+1}+q^{6 k_l n+4k_s+2}-q^{8 k_l}+\\
	  (q^{2 k_s+1}+1)  (q^{2 k_l (n+3)}-q^{4 k_l n+2 k_l+2 k_s+1})+\\
	  (q+1)  (q^{2 k_l}+1)  (q^{4 k_l n+2 k_l+3k_s+1}-q^{2 k_l (n+2)+ k_s}))
\end{eqnarray}

For $C_n$ algebras the same normalization is used, with the square of the long root $4$. Then $F_X$ writes as

\begin{eqnarray}
	F_{C_n}= \frac{A_{C_n}}{B_{C_n}} \\
	B_{C_n}=(q^2-1) (q^{2 k_s}-1) \\
	A_{C_n}=	(q^{k_s n}-1) q^{-2 k_l-k_s (2 n+1)} \times \\
	(q^{2 k_l+2 k_s (n+1)+1}-q^{2 k_l+k_s (n+2)+1}-q^{2 k_l+k_s (n+3)+1}+q^{2 (k_l+k_s n+k_s+1)}+\\
	q^{2 k_l+2 k_s n+k_s+1}-q^{4 k_l+2 k_s n+k_s+1}+q^{4 k_l+3 k_s n+1}-q^{2 k_l+3 k_s n+k_s+2}+\\
	q^{4 k_l+3 k_s n+k_s+2}+q^{4 k_l+3 k_s n+2}-q^{2 k_l+k_s (n+2)}+q^{2 k_l+3 k_s}+\\
	q^{k_s (n+3)+1}-q^{4 k_s+1}-q^{3 k_s}-q^{4 k_s})
\end{eqnarray}

For $F_4$, with the same normalization, we have 

\begin{eqnarray}
	F_{F_4}= \frac{A_{F_4}}{B_{F_4}} \\
	B_{F_4}=(q^2-1)   \\
	A_{F_4}=q^{-2 (5 k_l+3 k_s)} (q^{2 (2 k_l+k_s)}+1) (q^{6 k_l+6 k_s+1}-1) \times \\
	 (q^{4 k_l+k_s+1}+q^{6 k_l+3 k_s+1}-q^{4 k_l+4 k_s+1}+q^{10 k_l+4 k_s+1}+\\
	 q^{3 (2 k_l+k_s)}+q^{4 k_l+k_s}-q^{6 k_l}+1) 
\end{eqnarray}

For $G_2$ we use the normalization corresponding to the square of the long root to be equal to $6$. The corresponding $F_{G_2}$ function is

\begin{eqnarray} \label{FXG2}
	F_{G_2}= \frac{A_{G_2}}{B_{G_2}} \\
	B_{G_2}=q^3-1   \\
	A_{G_2}=q^{-3 (2 k_l+k_s)} (q^{3 k_l+3 k_s+1}-1) \times \\
	(q^{3 k_l+k_s}+q^{2 (3 k_l+k_s)}+q^{3 k_l+k_s+1}+q^{3 k_l+k_s+2}+q^{6 k_l+2 k_s+1}+\\
	q^{6 k_l+2 k_s+2}-q^{3 k_l+3 k_s+2}+q^{9 k_l+3 k_s+2}-q^{6 k_l}+1)
\end{eqnarray}

\section{Conclusion and outlook}
One can investigate the formulae above in various limits. E.g. a reasonable limit is $k_s=0$. 
In that case only the long roots contribute in formula (\ref{FX}). 
On the other hand, the long roots of a non simply laced algebras together with the Cartan subalgebra constitute some subalgebra $L_X$ of the original algebra  $X$. 

So, one has the following equality:

\begin{eqnarray}\label{ccc}
	F_X(x,k_l,0)=F_{L_X}(cx,k_l)
\end{eqnarray}
where in the r.h.s. we have functions for single-refined algebras, given in \cite{AM21}. The rescaling constant $c$ may appear due to different normalization of the long roots in
 algebras $X$ and $L_X$. 
 
 For example, for $G_2$ algebra one has
 
 \begin{eqnarray}
 	L_{G_2}=A_2; 
 \end{eqnarray}
 
  Then from (\ref{FXG2}) we have

\begin{eqnarray}
	F_{G_2}(x,k_l,0)=\frac{q^{-6 k_l} \left(q^{3 k_l}+1\right) \left(q^{9 k_l+3}-1\right)}{q^3-1}
\end{eqnarray}
and, on the other hand, for single-refined case one has \cite{AM21}

\begin{eqnarray}\label{FA2}
F_{A_2}(x,y)=	\frac{q^{-2y} \left(q^{y}+1\right) \left(q^{3y+1}-1\right)}{q-1}
\end{eqnarray}

These two formulae coincide after identification $k_l=y$ and redefinition of $q=\exp (x)$ (rescaling of $x$) in the second one according to the 

\begin{eqnarray}
	q \rightarrow q^{3}
\end{eqnarray}
i.e. in (\ref{ccc}) $c=3$. 
This redefinition is required since the square of the long roots in the $F_{G_2}(x,k_l,k_s)$ is 6, and in $F_{A_2}(x,y)$ it is 2. Altogether, we see that in the $k_s=0$ limit our 
formulae agree with the previously known ones.

This limit $k_s=0$ resembles the Nekrasov-Shatashvili (NS) \cite{NS09} limit in the refined Chern-Simons theories. Of course, the true NS limit is $k_s=k_l \rightarrow 0$. 
In our case an interesting new possibility is the limit when both $k_s$ and $k_l$ are tending to zero with $k_s/k_l$ fixed. 
Then one would have an additional deformation parameter $k_s/k_l$ and the question would be what kind of (deformed) quantum integrable systems would appear in this limit. 

The other important direction for a further investigation is the following. 
The present work generalizes the results obtained in \cite{AM21} where the partition function for single-refined Chern-Simons theory was presented for all gauge algebras and
was transformed into a duality-ready form (universal-like representation). The next step was taken in \cite{AM22} where the duality with the refined topological 
strings was established (for the first time in case of the non simply laced algebras). 
The procedure for carrying out that establishment included the transformation of the partition function into the form of product/ratio of multiple sine functions, as well as the 
perturbative approximation of the multiple sines which yielded a partition function for the topological strings of a Gopakumar-Vafa type. 
Let us look at the simplest case as an example. The refined $A_{N-1}$  Chern-Simons theory partition function can be transformed into

\begin{equation} 
	\begin{split}
		Z_A(a,y,\delta) &= 
		\sqrt{\frac{\delta}{ya}} \frac{S_3(1+ya|1,y,\delta)}{S_3(y|1,y,\delta)}\,,
	\end{split}
\end{equation}
with $a=N$. A similar form of the partition function can be obtained for other gauge algebras \cite{AM22}, but it requires more complex computations.

One can implement similar calculations to those in \cite{AM22} for the two-fold refined case. However, the situation with the double refined theories is more complicated in two respects. First, due to second refinement parameter the similar transformations of the partition functions are much more complicated. Second, there are no known two-fold refined 
topological strings to compare our formulae with.

\section{Acknowledgments.}
MA and RM are partially supported by the Science Committee of the Ministry of Science 
and Education of the Republic of Armenia under contract 21AG-1C060. 
The work of MA is partially supported by ANSEF grant PS-mathph-2697.

\end{document}